\documentclass[aps,prc,showpacs,twocolumn,letter]{revtex4}

\usepackage{graphicx,epsfig}

\begin{document}

\title{Longitudinal correlation of the triangular flow event plane
in a hybrid approach with hadron and parton cascade initial conditions}

\author{Hannah Petersen}
\affiliation{Department of Physics, Duke University, Durham, North Carolina
27708-0305, United States}

\author{Carsten Greiner}
\affiliation{Institut f\"ur theoretische Physik,  Johann Wolfgang
Goethe-Universit\"at, Max-von-Laue Strasse 1,
 D-60438 Frankfurt am Main, Germany}

\author{Vivek Bhattacharya}
\affiliation{Department of Physics, Duke University, Durham, North Carolina
27708-0305, United States}

\author{Steffen A. Bass}
\affiliation{Department of Physics, Duke University, Durham, North Carolina
27708-0305, United States}


\begin{abstract}
The longitudinal long-range correlations of the triangular flow event plane
angles are
calculated in a Boltzmann + hydrodynamics hybrid approach. The potential to
disentangle different energy deposition scenarios is explored by utilizing two
different transport approaches for the early non-equilibrium evolution. In the
hadronic transport approach the particle production in high energy heavy ion
reactions is mainly governed by string excitation and fragmentation processes
which are absent in the parton cascade approach. We find that in both approaches
the initial state shows a strong longitudinal correlation of the event plane
angles which is
diluted but still persists in the final state momentum space distributions of
the produced particles. A ridge-like structure can also be caused by
near-collinear
gluon radiation in a parton cascade approach and does not necessarily prove
longitudinal flux tubes in the initial state. 
\end{abstract}

\keywords{Relativistic Heavy-ion collisions, Monte Carlo simulations,
Hydrodynamic models}

\pacs{25.75.-q,24.10.Lx,24.10.Nz}

\maketitle

Long- range rapidity correlations have been measured in heavy ion reactions at
the Relativistic Heavy Ion Collider (RHIC)
\cite{Adams:2005ph,Adams:2004pa,Putschke:2007mi,:2009qa,Alver:2009id}. In
two-particle correlations with a high $p_T$ trigger hadron in the
two-dimensional
$\Delta \eta-\Delta \phi$ plane a ridge-like structure around $\Delta \phi=0$
that extends over several units in pseudorapidity difference has been observed.
Later on, the same phenomenon was also found in untriggered correlation
functions and more detailed measurements with respect to particle composition
and effective temperatures have been performed that indicate the
'bulk'-like properties of the ridge. Therefore, the original interpretation as a
jet-medium effect \cite{Majumder:2006wi,Shuryak:2007fu,Wong:2008yh} has been
questioned and other explanations as e.g. Color-Glass-Condensate (CGC)
inspired flux tube structures that get boosted by radial flow have become more
favored
\cite{Dumitru:2008wn,Gavin:2008ev,Dusling:2009ni,Moschelli:2009tg,
Moschelli:2009bk}. 

Recently, initial state fluctuations have been proposed as the major source for
many of the structures that appear in two-particle correlations
\cite{Alver:2010gr,Luzum:2010sp,Hama:2009vu,Sorensen:2011hm}. Especially the
third Fourier coefficient of the azimuthal distribution of the final state
hadrons
in momentum space, namely triangular flow, is studied with great interest
\cite{Petersen:2010cw,Qin:2010pf,Ma:2010dv}. This
new flow observable is directly related to the fluctuations in the initial state
and is absent in hydrodynamic calculations assuming smooth initial state
profiles. Triangular flow is therefore independent of the collision centrality
and very sensitive to the viscosity of the produced matter
\cite{Alver:2010dn,Schenke:2010rr,Schenke:2011tv,Teaney:2010vd}.

Even though a lot of progress has been made in the study of initial conditions,
in particular regarding their fluctuations, it has been 
rather difficult to find an observable that
is directly related to the mechanism of the initial energy deposition.
Long-range
rapidity correlations seem to be perfectly suited for this purpose, since they
need to be build up very early during the evolution of the reaction due to 
causality. In order to study the effect of initial state fluctuations one needs
to simulate the whole dynamical evolution event-by-event, which is
computationally
very expensive and makes it difficult to study two-dimensional 2-particle
correlations
in detail. 

In this paper we propose a new observable that directly quantifies the
longitudinal
long-range correlation of triangular flow and explore it utilizing a 
state-of-the-art (3+1)d Boltzmann+hydrodynamics approach
\cite{Petersen:2008dd}. 
The initial conditions are either generated
using hadron-string
dynamics from the Ultra-relativistic Quantum Molecular Dynamics (UrQMD) approach
\cite{Bass:1998ca,Bleicher:1999xi}
or by using a parton cascade approach (PCM) \cite{Geiger:1994he,Bass:2002fh}. 
In both scenarios the longitudinal
correlations of the event plane angles are calculated in a way that is also
accessible in experiment. 

\begin{figure}[ht]
\resizebox{0.5\textwidth}{!}{ \centering
\includegraphics{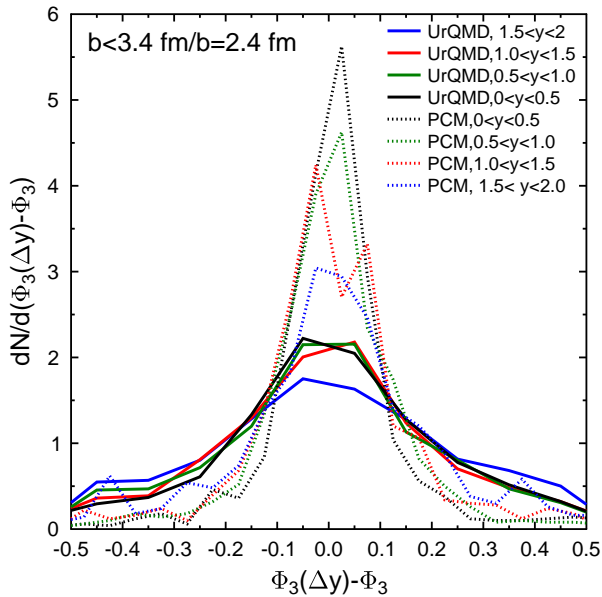}
}
\caption{(Color online) Distribution of the differences of the coordinate space
event-plane angles $\Phi_3$ in different rapidity slices in UrQMD (full
line) and PCM (dotted line) initial conditions for central ($b<3.4$ fm/$b=2.4$
fm) Au+Au collisions at $\sqrt{s_{\rm NN}}=200$ GeV.} 
\label{fig_urqmdpcm_inicen}
\end{figure}

\begin{figure}[ht]
\resizebox{0.5\textwidth}{!}{ \centering
\includegraphics{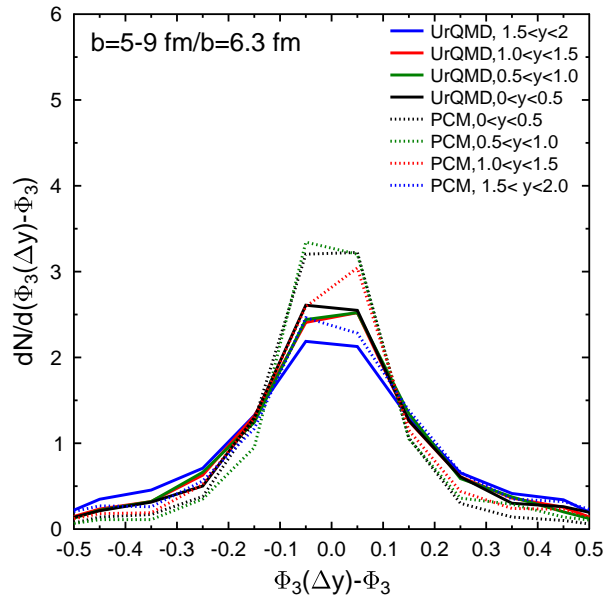}
}
\caption{(Color online) Distribution of the differences of the coordinate space
event-plane angles $\Phi_3$ in different rapidity slices in UrQMD (full
line) and PCM (dotted line) initial conditions for mid-central ($b=5-9$
fm/$b=6.3$ fm) Au+Au collisions at $\sqrt{s_{\rm NN}}=200$ GeV.} 
\label{fig_urqmdpcm_inimid}
\end{figure}

Let us first describe the two different approaches for the initial conditions in
more detail. For the hadron-string dynamics, the initial binary nucleon-nucleon
collisions are modeled in UrQMD
\cite{Bass:1998ca,Bleicher:1999xi,Petersen:2008kb} following the Lund model of
nucleon-nucleon
reactions \cite{Andersson:1983ia} involving color flux tubes excitation
and fragmentation processes that provide long range rapidity correlations and
fluctuations in the energy deposition in
the transverse plane. The other option is to decompose the incoming nuclei
according to parton distribution functions into quarks and gluons and use a
parton cascade approach to simulate the early non-equilibrium evolution
\cite{Geiger:1991nj,Geiger:1994he,Geiger:1997pf,Bass:2002fh}. In this
case, the cross-sections for 2$\rightarrow$2 collisions and 1$\rightarrow$2
splittings  are
given by perturbative QCD calculations. We have chosen this approach, since at
first glance 
it does not contain any obvious
mechanism to generate long-range correlations and thus provides
a baseline to compare to. However, as we shall later see, 
the presence of radiated parton showers generated after the initial hard
parton-parton scatterings is capable of generating long range rapidity
correlations.

For Au+Au collisions at the highest
RHIC energies the
starting time for the hydrodynamic
evolution has been chosen to be $t_{\rm start}=0.5$ fm in order to fit the final
state
pion multiplicity at midrapidity. Only the
matter around midrapidity ($|y|<2$) is considered to be locally thermalized and
takes part in the ideal hydrodynamic
evolution. To map the point
particles from the hadron/parton cascade initial state to energy, momentum and
net baryon
density distributions each particle is
represented by a three-dimensional Gaussian distribution
\cite{Steinheimer:2007iy}. The width of the Gaussian distribution has been
chosen to be $\sigma=1$ fm for UrQMD and $\sigma=0.5$ fm for the PCM. 

\begin{figure}[ht]
\resizebox{0.5\textwidth}{!}{ \centering
\includegraphics{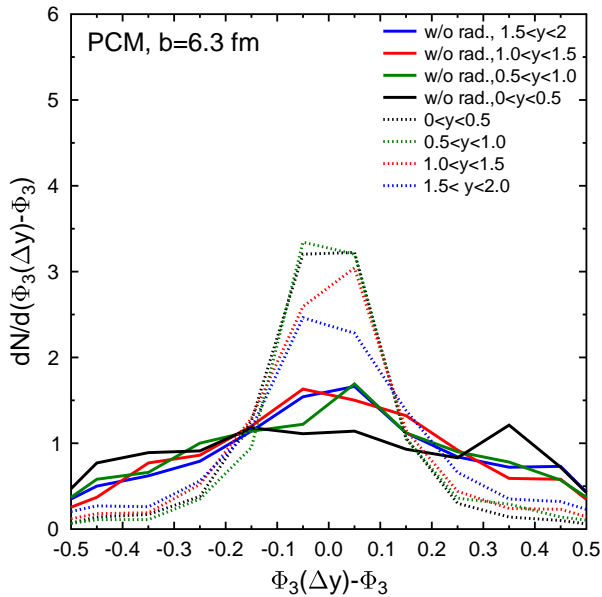}
}
\caption{(Color online) Distribution of the differences of the coordinate space
event-plane angles $\Phi_3$ in different rapidity slices from the parton
cascade in the default scenario (dotted 
line) and without gluon radiation (full line) initial conditions for mid-central
($b=6.3$ fm) Au+Au collisions at $\sqrt{s_{\rm NN}}=200$ GeV.} 
\label{fig_rad_pcm_phi3}
\end{figure}

The ideal hydrodynamic evolution \cite{Rischke:1995ir,Rischke:1995mt} for the
hot and dense stage of the collision
translates the initial fluctuations in the transverse energy density to momentum
space distributions. A hadron gas
equation of state \cite{Zschiesche:2002zr} has been used for the calculation
with UrQMD initial conditions. 

The transition from the hydrodynamic evolution to the transport approach when
the matter is diluted in the late stage
is treated as a gradual transition on an approximated constant proper time 
hyper-surface (see \cite{Li:2008qm} for details). For the hadronic calculation
an energy density of 713 MeV/fm$^3$ has been chosen as a transition criterion.
Late stage hadronic rescattering and
resonance decays are taken into account
in the hadronic cascade.

The above event-by-event setup includes all the main ingredients that are
necessary for the build up of triangular flow \cite{Petersen:2010cw}. Since the
complete final state particle distributions are
calculated, an analysis similar to those applied by experimentalists is used.

\begin{figure}[ht]
\resizebox{0.5\textwidth}{!}{ \centering
\includegraphics{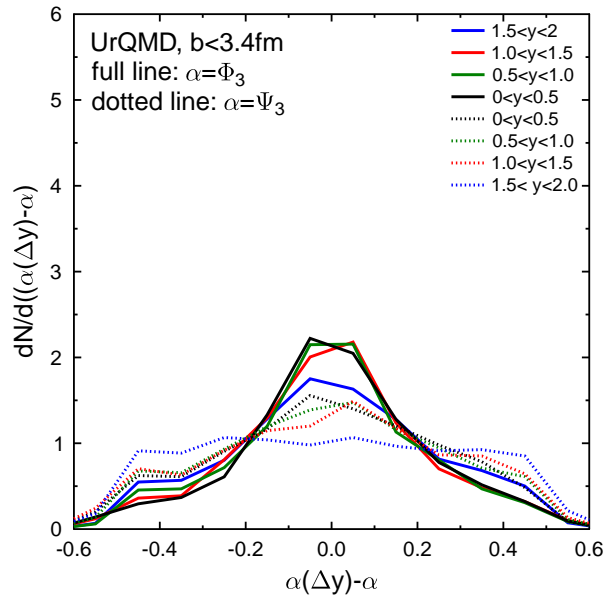}
}
\caption{(Color online) Distribution of the differences of the final momentum
space (dotted line) vs initial coordinate space (full line) event-plane angles
$\Phi_3$ in different rapidity slices in the hybrid approach based on
UrQMD initial conditions for central ($b<3.4$ fm) Au+Au collisions at
$\sqrt{s_{\rm NN}}=200$ GeV.} 
\label{fig_urqmdfincen}
\end{figure}

\begin{figure}[ht]
\resizebox{0.5\textwidth}{!}{ \centering
\includegraphics{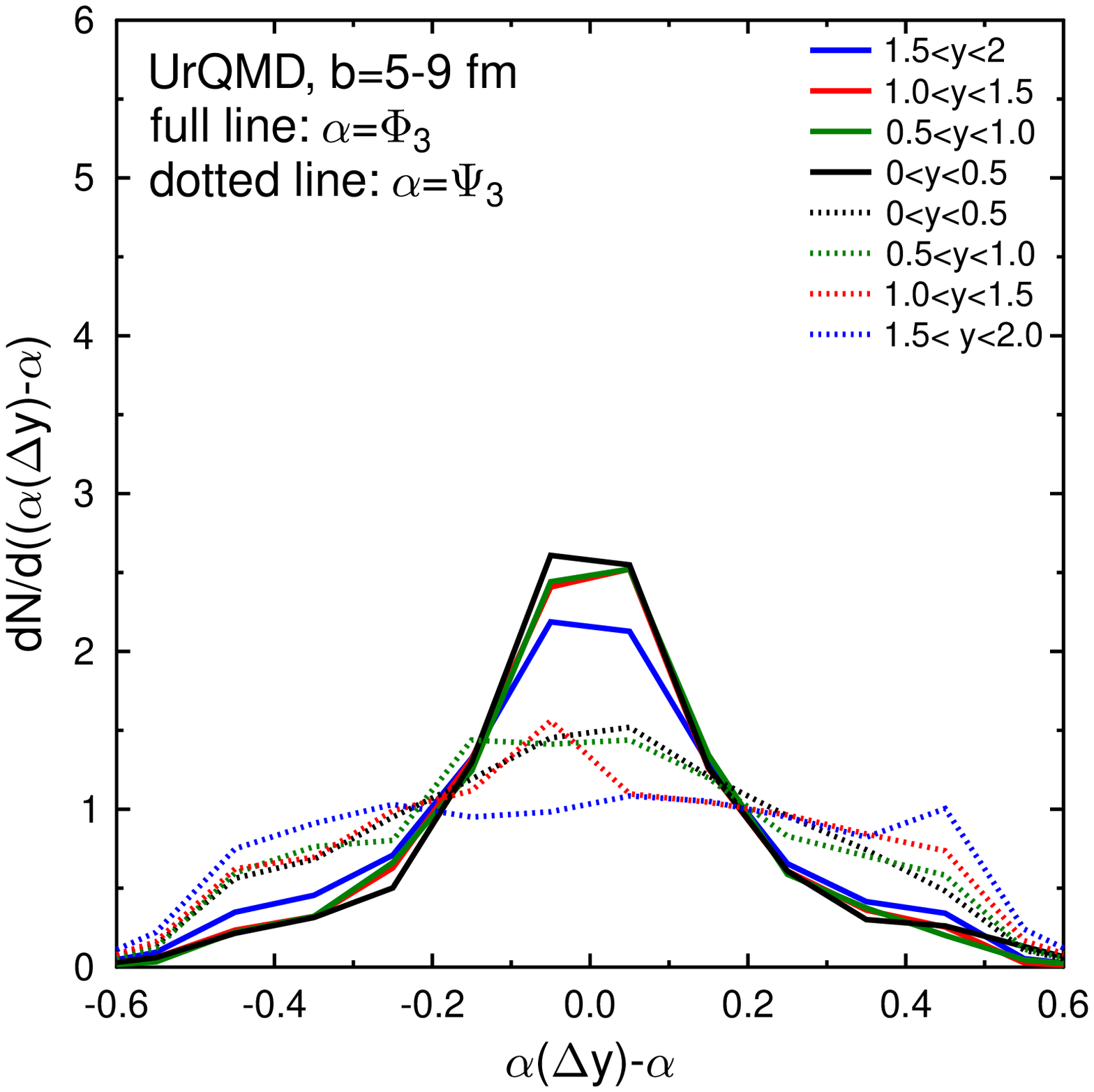}
}
\caption{(Color online) Distribution of the differences of the final momentum
space (dotted line) vs coordinate space (full line) event-plane angles $\Phi_3$
in different rapidity slices in the hybrid approach based on UrQMD initial
conditions for mid-central ($b=5-9$ fm) Au+Au collisions at
$\sqrt{s_{\rm NN}}=200$ GeV.} 
\label{fig_urqmdfinmid}
\end{figure}

In \cite{Petersen:2010cw} the pseudorapidity dependence of triangular flow in
the hybrid approach explained above has been published. The flatness over a
broad range of pseudorapidities indicates long-range correlations, but does not
actually prove that correlation. The preferred axis of the triangular shape
could
randomly fluctuate from one rapidity bin to the other and one would still observe
a flat rapidity dependence as long as the magnitude of the third Fourier
coefficient is the same in every bin. To really study the long-range correlation
between the 'hot spots' we propose to investigate the triangular flow event
plane in different rapidity bins separately and calculate the correlation
between these. 

In the initial state coordinate space distributions, one can define the event
plane angles for different harmonics (we will concentrate on $n=2$ and $n=3$ in
this
analysis) in the following way: 

\begin{equation}
\Phi_n=\frac{1}{n} {\rm arctan}\frac{\langle r^n \sin(n\phi)\rangle}{\langle r^n
\cos(n\phi)\rangle} \quad 
\end{equation}

\noindent where $r$ and $\phi$ are the polar coordinates of the particles in the
center of
mass frame of the collision. For elliptic flow this angle is the one that
defines the so called
participant
plane. With the used conventions $\Phi_n$ is defined in the region between
$-\pi/n$ and $+\pi/n$. To explore the longitudinal correlations, the event plane
angle is calculated once in the whole rapidity range from $-2<y<2$ and
then in different bins of width $|\Delta y|<0.5$ separately. The
distributions of
the difference between those two angles in different rapidity slices is
shown in Fig. \ref{fig_urqmdpcm_inicen}. We have checked that the results are
symmetric around midrapidity and the results are qualitatively unchanged if the
bin size is varied from $|\Delta y|<0.25-1.0$. Furthermore, the result
without any
additional physics mechanism that introduces a long-range correlation is a flat
distribution (corresponding to a $\delta$-function in the dN/d$\Delta y$
distribution) which we can reproduce by sampling particles from different events
together and apply the exact same correlation analysis, similar to a mixed event
technique.

In Fig. \ref{fig_urqmdpcm_inicen} and \ref{fig_urqmdpcm_inimid} the correlation
of the event plane angles is shown for two different centralities and the two
different initial state transport approaches. In both cases, there is a strong
correlation visible that is largest at midrapidity and smaller at the most
forward $y$ bin. In the parton cascade approach the longitudinal correlation
can be attributed to the emission of parton showers that appear to be correlated
to
the following hard collision and are emitted along the beam axis, whereas in the
hadronic transport approach the string excitation and fragmentation mechanism
offers the most plausible explanation. 

Fig. \ref{fig_rad_pcm_phi3} shows a comparison for the parton cascade initial
conditions with and without time-like branchings, which initiate the parton
showers. Without gluon emission after the hard collision, the longitudinal
correlation of the triangular flow event plane angle in the initial state is
much smaller (if existent at all). That proves that the initial long-range
correlation in the parton cascade is mainly caused by the initial gluon
radiation. 

To investigate if the initial long-range correlation survives the hydrodynamic
expansion the same analysis can be performed for the final state momentum space
event plane angles, 

\begin{equation}
\label{eqn_defpsi}
\Psi_n=\frac{1}{n} {\rm arctan}\frac{\langle p_T \sin(n\phi_p)\rangle}{\langle
p_T \cos(n\phi_p)\rangle} \quad ,
\end{equation}
where $(p_T,\phi_p)$ are polar coordinates in momentum space. In Figs.
\ref{fig_urqmdfincen} and \ref{fig_urqmdfinmid} the longitudinal correlation of
the final state event plane angles $\Psi_3$ is shown in comparison to the
previously presented distribution for $\Phi_3$. The correlation gets
significantly
diluted during the ideal hydrodynamic expansion, but it persists also in the
final state particle distribution with the exception of the most
forward/backward rapidity bin that is accessible in this calculation. We have
checked that the same result qualitatively is obtained by employing the parton
cascade initial conditions. The final state event plane angles $\Psi_3$ in
different rapidity bins could be measured in
experiment to investigate the longitudinal correlations of triangular flow and
to legitimate the assumption that the event plane angle does not change as a
function of rapidity.

\begin{figure}[h]
\resizebox{0.5\textwidth}{!}{ \centering
\includegraphics{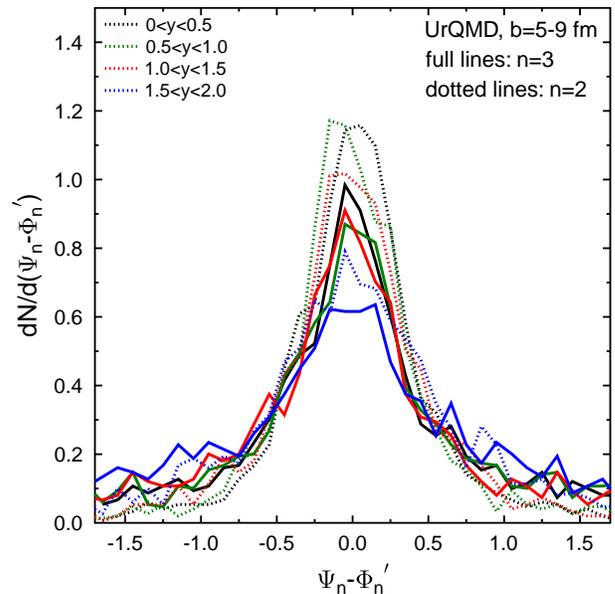}
}
\caption{(Color online) Event-by-event correlation between initial and final
state event plane angles for elliptic (dotted line) and triangular flow (full
line) in different rapidity slices in the hybrid approach based on UrQMD
initial conditions for mid-central ($b=5-9$ fm) Au+Au collisions at
$\sqrt{s_{\rm NN}}=200$ GeV.} 
\label{fig_urqmdcorr_eta}
\end{figure}

Another way to study the initial and final state correlation of the event plane
angles is shown in Fig. \ref{fig_urqmdcorr_eta}. Here, the differences between
the two angles in different rapidity bins are shown for mid-central
collisions in the hybrid approach based on UrQMD initial conditions. As a
comparison the correlation for elliptic flow at is also shown. For
the three rapidity bins between 0 and 1.5 the correlation is clearly
visible for triangular flow as well, whereas it is less pronounced in the
most forward/backward bin. This confirms the conclusion of the Fig.
\ref{fig_urqmdfinmid} that the long-range correlation only survives over three
units in rapidity around $y=0$. 

In summary, we have proposed a new way to investigate the long-range
correlations of triangular flow in heavy ion collisions. A hadron and a parton
cascade approach both lead to sizable long-range correlation in the initial
state.
These correlations are translated to the final state hadron distributions, but
are
significantly diluted and smoothed out during the final hadronic rescattering. 
Measuring the triangular flow event plane angles in different rapidity
bins serves as a pre-requisite to legitimate the assumption of a constant event
plane over rapidity as it is commonly assumed by experimental groups. 
The second conclusion is that one needs to take into account a realistic
dynamical evolution (e.g. \cite{Xu:2004mz,Xu:2007aa}) to prove that the
long-range
correlations that might be initially established also survive to the final state
particle distributions instead of simple parametrizations.

\section*{Acknowledgements}
\label{ack} We are grateful to the Open Science Grid for the computing
resources. The authors thank Dirk Rischke for
providing the 1 fluid hydrodynamics code. This work was
supported by the Hessian LOEWE initiative through the Helmholtz International
Center for FAIR (HIC for FAIR). H.P. acknowledges a Feodor Lynen
fellowship of the Alexander von Humboldt
foundation. This work was supported in part by U.S. department of Energy grant
DE-FG02-05ER41367 and NSF grant PHY-09-41373. The authors thank G. Qin, R.
Snellings, N. Xu and Z. Xu for fruitful discussions.


\end{document}